\documentclass[%
 11pt,
  a4paper,
]{article}

\usepackage{GRFessay}
\usepackage{slashed}
\usepackage{mathrsfs}

 \title{Geometry creates inertia}
  \author{Amitabha Lahiri}
  \emailAdd{amitabha@bose.res.in }
  \affiliation{S. N. Bose National Centre for Basic Sciences\\
  JD Block, Sector III, Salt Lake, Kolkata - 700106}

\abstract{
The dynamics of fermions in curved spacetime is governed by a spin connection, a part of which is contorsion, an auxiliary field independent of the metric, without dynamics but fully expressible in terms of the axial current density of fermions. Its effect is the appearance of a quartic interaction involving all fermions. Contorsion can couple to left and right-handed fermions with different strengths, leading to an effective mass for fermions propagating on a background containing fermionic matter. 

\vskip 0.5in

\centerline{\em Essay written for the Gravity Research Foundation 2020 Awards for Essays on Gravitation}

\medskip
\centerline{\em Submitted on 30th March, 2020}

\bigskip
\centerline{\em Selected for Honorable Mention}
}
	 
\begin{document}

\maketitle


``Inertia originates in a kind of interaction between bodies", wrote Einstein in a letter to Ernst Mach~\cite{Einstein:1913a} when he was working hard to find a relativistic theory of gravitation, an intense effort that would culminate two years later in the General Theory of Relativity. Einstein's comment is now interpreted in the context of his theory as ``Space tells matter how to move; matter tells space how to curve''~\cite{Misner:1974qy}. In this essay I argue that inertia of matter and geometry of spacetime are even more closely intertwined -- spacetime geometry creates inertia for elementary particles of matter by inducing quantum interactions with all matter in the immediate vicinity.


The Dirac equation governing the dynamics of fermions was not yet known when Einstein wrote down his theory of gravitation. Thus it was not until much later that it was realized that the Levi-Civita connection does not completely specify how the covariant derivative operator acts on spinors. In curved spacetime, the Dirac matrices  $\gamma_a$ are defined on an ``internal" flat space, isomorphic to the tangent space at each point.``Spacetime" $\gamma$ matrices are then defined by introducing tetrad fields as $\gamma_\mu =  e^a_\mu \gamma_a\,.$ Then it is convenient to describe gravity coupled to both bosonic and fermionic fields in terms of tetrads and a spin connection $A^{ab}_\mu$ in a formulation that can thought of as a gauge theory of gravity~\cite{Cartan1922,Kibble:1961ba,Sciama:1964wt,Hehl:1976kj}, with $A^{ab}_\mu$ being treated as a SO(3,1) gauge field {\em a priori} independent of the tetrads~\footnote{
	{We will use Greek indices for spacetime and Latin indices for the internal space, the tetrads relating the metric $g_{\mu\nu}$  of spacetime with the Minkowski metric $\eta_{ab}\, (-+++)$ of the local tangent space through the relations $\eta_{ab}e^I_\mu e^J_\nu = g_{\mu\nu} $\,. }
	
}.

If gravity is minimally coupled only to bosonic fields, it can be written as a theory using tetrads and spin connection but without an explicit metric. It is a first order formulation of Einstein gravity and is fully equivalent to it. The gravitational Lagrangian is written using these fields as $\frac{1}{2\kappa} |e|  F^{ab}_{\mu\nu}\, e^\mu_a\, e^\nu_b $\,, where $F$ is the field strength of the gauge field $A$\,, $e$ is the determinant of the matrix $e^a_\mu$\,, and $\kappa = 8\pi G$\,. If we include only the known bosonic fields which couple minimally to gravity, the unique solution for $A_\mu^{ab}$ is $\omega_\mu^{ab}$\,, which corresponds to the unique torsion-free connection in the metric formulation. Then the equation of motion for the tetrads become identical with the usual Einstein equation.

This is not the complete story when fermions are present. 
In this case,  $A_\mu$ has another component $\Lambda_\mu$ called {\em contorsion}, $A^{ab}_\mu = \omega^{ab}_\mu  + \Lambda^{ab}_\mu\,$. The action, for gravity and one species of fermions with all other fields suppressed, is
\begin{equation}\notag
S = \int |e| d^4x \left[ \frac{1}{2\kappa} F^{ab}_{\mu\nu} e^\mu_a e^\nu_b 
+ \frac{i}{2} \left( \bar{\psi}\slashed{D}\psi - (\bar{\psi}\slashed{D}\psi)^\dagger \right)
+im\bar\psi\psi  \right]\,. 
\end{equation}
Then $\Lambda_\mu$ turns out to be an auxiliary field whose equation of motion  has the unique solution $\Lambda^{ab}_\mu = \frac{\kappa}{8}\epsilon^{ab}_{\phantom{IJ}cd}\,  e^c_\mu \bar{\psi}\gamma^d\gamma^5\psi\,.$ This is for only one species of fermion. Since gravity couples to all fields, $\Lambda$ will couple to, and get contributions from, all species of fermions. In the absence of fermions $\Lambda$ vanishes, irrespective of any bosonic fields present as long as they are minimally coupled to gravity.

Invariance under local Lorentz transfomations means that $\omega$ transforms inhomogeneously under them, while $\Lambda$ transforms homogeneously.  Thus the coupling of $\Lambda$ to fermions is not like that of a gauge field but like Yukawa coupling -- the coupling strength can be freely set by hand.
But unlike a scalar field, $\Lambda$ being a vector can couple chirally to fermions -- after all, geometry dictates that it couple to left-handed neutrinos irrespective of whether or not there are right-handed neutrinos in the universe. So in general fermions of different species and chirality could couple to $\Lambda$ with different coupling strengths, analogous to the Yukawa coupling of fermions to a scalar field.

Thus I can now write the most general action for fermions on a curved background and from there find the corresponding expression for the contorsion~\cite{Chakrabarty:2019cau},  
\begin{eqnarray}\notag
 \Lambda_\mu^{ab}  = 
 && \frac{\kappa}{8} \epsilon^{ab}_{\phantom{IJ}cd} \, e_\mu^c \sum_f \left(\lambda_{fL}\bar{f}_{L} \gamma^d \gamma^5 f_{L} + \lambda_{fR}\bar{f}_{R} \gamma^d \gamma^5  f_{R}\right)\,
\end{eqnarray}
where fermion fields are denoted by $f$, with left- and right-handed components marked by subscripts $L$ and $R$\,. 
In the metric picture $\Lambda$ corresponds to a torsion $C^\alpha_{\phantom{\alpha}\mu\nu} = \Lambda^{ab}_{[\mu} e_{\nu]b} e^\alpha_a $ 
which is totally antisymmetric and therefore does not affect geodesics. Thus all particles fall at the same rate  in a gravitational field and the principle of equivalence is not violated by choosing different coupling constants for different particles or different chiralities. 

Since $\Lambda$ is an auxiliary field and must always equal its on-shell value, it is safe to insert the solution of $\Lambda$ back into the action. If I write $\Lambda$ using the vector and axial currents, the action of fermion fields in curved spacetime has the form
\begin{eqnarray}\notag
S = \int |e| d^4x &&\left[
\frac{i}{2}\sum_f \left( \bar{f}\slashed{D} {f} - (\bar{f}\slashed{D} {f})^\dagger  + 2 m_f\bar{f} f \right)
-\frac{3\kappa}{16}\left(\sum_f \left( \lambda_{fV} \bar{f} \gamma^d {f} + \lambda_{fA} \bar{f} \gamma^d\gamma^5 {f} \right)\right)^2 \right]\,.\qquad
\end{eqnarray}
The last term represents a four-fermion interaction which is uniquely gravitational in nature. On a flat background $\omega^{ab}_\mu$ vanishes, but no matter how close the spacetime is to flatness, this quartic term must be present in the action as long as Newton's constant is nonvanishing. The only ways this term can be absent from the action are if gravity is turned off ($\kappa = 0$), or if the quartic couplings $\lambda_f$ are set to zero by hand. The curvature of the background has no effect on the quartic term. 

Fermions gain an effective mass from this term via quantum interactions with other fermions when passing through matter. To see this, consider one species of neutrino and assume that it is massless in vacuum and also purely left-handed. Then the quartic interaction between the neutrino and other fermions is
\begin{eqnarray}\notag
{\mathscr L}_{\bar{\nu}\nu\bar{f}f}
& = & \frac{3\kappa}{8} \lambda_{\nu}(\bar{\nu}\gamma^\mu\nu)
{\sum_{ f\neq \nu}}  \left(\lambda_{fV}\bar{f} \gamma_\mu {f}
+ \lambda_{fA}\bar{f} \gamma_\mu\gamma^5 {f}\right)\,.
\end{eqnarray}
Most of the known matter in the universe is in the form of electrons, protons, and neutrons. For neutrinos passing through such matter, the elastic forward scattering amplitude is~\cite{Wolfenstein:1977ue}
\begin{equation}\notag
{\cal M} = \frac{3\kappa}{8}  \lambda_{\nu} \bar{\nu}\gamma_\mu\nu \sum_{f=e,p,n}   \left\langle\left(\lambda_{fV}\bar{f} \gamma^\mu {f}
+ \lambda_{fA}\bar{f} \gamma^\mu\gamma^5 {f}\right) \right\rangle\,,
\end{equation}
where the average is taken over the background. For the axial current, the average of the spatial components in the nonrelativistic limit is the spin, which for normal matter is negligible. The axial charge is also negligible. Similarly, the spatial components of the vector current average to the spatial momentum of the background, which can also be neglected.  

What we are left with is the average of the temporal component of the vector current of fermions, which is simply the number density of the fermions~\footnote{We are being a bit sloppy here -- the ``density" of the fermion field is the time component of $j^\mu \equiv e^\mu_I\bar\psi\gamma^I\psi$. In a 3+1 decomposition with $g_{00} = -\lambda^2$\,, we can write 
$j^0 = -\lambda^{-1}{\psi}^\dagger\psi$\,, with the volume measure $\lambda\sqrt{^3g}$\,, so we can refer to $f^\dagger f$ as the density and integrate it with the measure $\sqrt{^3g}$\,.},
$\langle \bar{f} \gamma^0 f\rangle = -\langle {f}^\dagger {f}\rangle = -n_f$\,. 
The contribution of the forward scattering amplitude to the effective Hamiltonian density is thus
\begin{equation}\notag
\delta {\mathscr H}_{\text{eff}} =  \lambda_{\nu} {\nu}^\dagger \nu\sum_{f=e,p,n}  \lambda_{f} n_f \,,
\end{equation}
where I have now dropped the subscript $V$\, and rescaled each $\lambda$ by $\sqrt{\frac{3\kappa}{8}}$\,. This term is an effective mass term for the neutrino, with $m_\nu = \lambda_{\nu}\rho$\,, where $\rho = \sum  \lambda_{f} n_f$\, is a weighted density of fermions. Since different $\lambda$'s are independent, the quartic term will also produce neutrino oscillations~\cite{Chakrabarty:2019cau}. 
 Not only neutrinos, but all fermions get a contribution to their masses in this manner. 

The $\lambda$'s can be bounded by experimental data such as anomalous dipole moments and neutrino masses, and are small enough that at low energies and normal matter densities, the only observable effects of the quartic term and the resulting ``geometrical inertia" will be on neutrinos. At higher densities such as in dense stars, the geometrical inertia of electrons and nucleons can become large, comparable to or larger than their masses in vacuum. So the actual mass of a dense star can be much larger than what is calculated from the electrons and baryons, thus affecting dark matter estimates. 
There are of course many other possible applications of this quartic interaction induced by geometry, especially in matter at high density as in collapsing stars or in the early universe. There is no symmetry which prevents the appearance of this term. 

{\em Thus irrespective of any other mechanism of mass generation, general covariance ensures that every fermion obeys a nonlinear Dirac equation, leading to a contribution to its inertia from the background density of other fermions.}


I conclude by making a couple of final remarks. The effective fermion action appears to be nonrenormalizable by power counting because of the quartic term. I have not worried about this issue here because the quartic couplings contain in them a factor of $\sqrt{\kappa}$ and thus the counterterms in curved spacetime will have to involve curvature. Therefore the question of renormalizability cannot be addressed without a theory of quantum gravity. 

The second point is about the size of the quartic term. Does the factor of $\sqrt{\kappa}$ which was absorbed in the $\lambda$'s make them too small? This question cannot be answered by pure logic and must be based on experimental observations. Unlike in the case of weak interactions, where the energy required to create $W$-boson pairs from the vacuum sets the scale of the four-fermion interaction, here the scale is not related to the quantum dynamics of $\Lambda$\,, which does not in fact have any dynamics. Therefore the coupling constants $\lambda$ are free and can be set only by comparison with experimental data, not from any theoretical argument.

\end{document}